\documentclass[aps,prx,twocolumn,groupedaddress, floatfix,longbibliography,byrevtex]{revtex4-2}

\usepackage{graphicx}
\usepackage[utf8]{inputenc}
\usepackage{amsmath,amsfonts,calc}
\usepackage{comment}
\usepackage[normalem]{ulem}
\usepackage[usenames,dvipsnames]{color}
\usepackage{soul}
\usepackage{graphicx}
\usepackage{scalerel}
\usepackage{mathtools}
\usepackage{stackengine,wasysym}
\usepackage[free-standing-units=true]{siunitx}

\begin{document}
\title{Threading an atom with light}

\author{Rodrigo G. Corti\~nas}
\email{rodrigo.cortinas@yale.edu}
\affiliation{Department of Applied Physics and Physics, Yale University, New Haven, CT 06520, USA}
\affiliation{Yale Quantum Institute, Yale University, New Haven, Connecticut 06520, USA}
\date{March 2023}
\begin{abstract}
We propose a ponderomotive trapping mechanism for circular Rydberg atoms that consists in threading the Rydberg orbital with a tightly focused Gaussian laser beam. The trap exhibits remarkable properties: it can be made effectively linear, extremely nonlinear, or have the same trapping frequency, depth, or nonlinearity for distant Rydberg states. The mechanism is capable of trapping efficiently highly excited Rydberg states with a single active electron, as well as ground state atoms using a single beam. Alternatively, the Rydberg atom can be trapped efficiently by the same beam used to excite it, relaxing optical constraints in well-shielded environments. 
\end{abstract}
\maketitle

\textit{Introduction--} Rydberg atoms, which are highly excited atoms with principal quantum number $n\gg1$, are interesting tools for quantum technologies and fundamental science \cite{saffman2010,adams2019}. They have long lifetimes and strong dipole moments originating from their enormous size \cite{gallagher2006}. For $n\geq 50$, for example, these atoms are larger than some viruses. These properties have been exploited to explore the principles of quantum mechanics \cite{haroche2006}, pioneer quantum computation and quantum simulation technologies \cite{haroche2013,browaeys2020,Cong2022}, and served as metrological tools \cite{osterwalder1999,sedlacek2012}.

Recently, laser trapping of Rydberg atoms has been demonstrated \cite{Cortinas2020,Graham2019, Barredo2020,Wilson2022} and this opened new technological perspectives at the single-atom level \cite{ravon2023_paper}. Rydberg atoms can now be trapped close to surfaces \cite{thiele2014,hermann2014,Morgan2020}, moved around at will, and used as sensitive single quantum sensors \cite{facon2016,dietsche2019}. As a result, they could serve as a stable ancilla to control bosonic quantum error correction codes in cavity quantum electrodynamics, and could be kept in interaction with other atoms for long times to perform quantum computations and simulations \cite{Teixeira2015,Nguyen2018,haroche2020,Cohen2021,Cong2022}.

Among Rydberg atoms, circular Rydberg atoms are particularly interesting \cite{Hulet1983,haroche2006} due to their exaggerated properties \cite{haroche2013,Nguyen2018}. These are atomic states of maximal angular and magnetic quantum numbers $l=m=n-1$, usually realized using alkali metal atoms with a single active electron. These atoms, having no optically allowed transitions, can be laser trapped by a ponderomotive potential \cite{landau1976,Dutta2000}, which means that they are low-intensity seekers. This constraint led the first experiments \cite{Cortinas2020} to use complex wavefronts, produced by spatial light modulators, with a dark spot surrounded by laser light. This puts significant constraints on the optical engineering required for trapping them, as well as requiring higher laser power needed to trap large $n>80$ states: these traps become shallower as the principal quantum number $n$ grows until they become anti-trapping \cite{Barredo2020}. The complex light intensity distribution typically has ``holes" which makes these traps laser-power inefficient. This is concerning when considering well-shielded cryogenic applications for quantum simulations \cite{Nguyen2018}, microwave-to-optical conversion of quantum information \cite{petrosyan2019,Vogt2019,liu2021,Kumar2023,Borowka2023}, or coupling to superconducting circuits \cite{hogan2012,haroche2020}, but especially so when considering trapping tens of thousands of individual Rydberg atoms for computation, and the ambition of high-fidelity Rydberg-based quantum gates \cite{Cohen2021}.

We propose here a power-efficient ponderomotive trapping mechanism that holds these atoms from within the electronic orbital and solely requires a tightly focused Gaussian beam. This trapping mechanism is original in that it depends on the shape and geometry of the atomic orbital. In its simplicity, it provides the flexibility to use the same beam to trap both the ground state and the circular Rydberg state of an alkali metal atom. Alternatively, a short wavelength beam, typically used for Rydberg excitation (see \cite{Cortinas2020} for a possible excitation scheme), can be used as a power-efficient Rydberg trapping beam. Both techniques relax optical power requirements and engineering constraints on the submicrometric alignment of the beams: this is hard in a vacuum chamber with little optical access, and especially in cryostats. 

Besides its fundamental interest as a new handle on the atomic orbital, the practical advantages mentioned above together with a set of properties we will derive below, make this thread trap interesting technologically. Among its properties, we will highlight at this point the trap's tunable nonlinearity. The high-nonlinearity mode could be used to observe nonlinear quantum dynamics of a single atom and, in particular, the generation of Schr\"odinger Kerr-cats \cite{yurke1986,haroche2006}. Note, that this deviation from classical motion has not been observed for a massive particle and has direct application to autonomous quantum error correction \cite{puri2017,grimm2020}. The opposite regime of the thread trap, the linear mode, is preferred for applications of future Rydberg quantum computation, simulation, and sensing. 

\begin{figure}[t]
    \centering
    \includegraphics[]{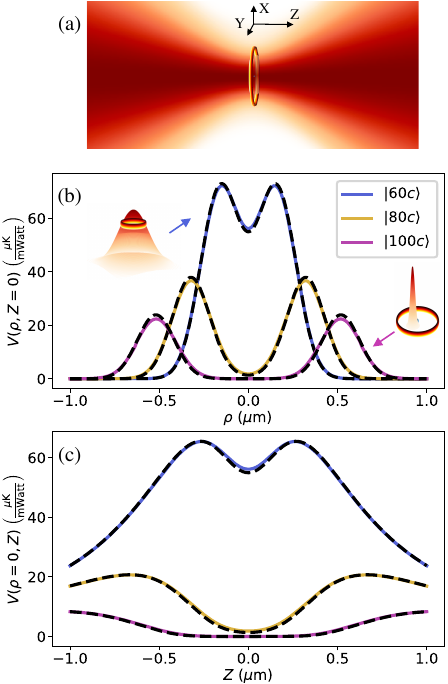}
    \caption{\textbf{Thread trap for circular Rydberg states $n=$ 60, 80, and 100.} (a) Sketch of the thread trap. (b) Radial trapping potential as a function of the distance between the nucleus and the beam's propagation axis $\rho$ at $Z=0$. The focal point of the beam is at the coordinate origin. The trapping laser is taken at a wavelength of $\lambda=420$~nm, which is a typical Rydberg excitation laser (see, for example, \cite{Cortinas2020}) and focused to a Gaussian waist of $w_0 = \lambda/2$. We show also a schematic of the Gaussian beam within the atomic orbital in two limiting cases. (c) Axial trapping potential as a function of the atomic position $Z$ along the beam's propagation axis and $\rho=0$. The solid lines correspond to the integration of the laser intensity over the 3D toroidal wavefunction $\psi_{nc}$. The dashed lines correspond to a curvilinear integration of the laser intensity along the semiclassical Bohr orbit. }
    \label{fig:80c_trap}
\end{figure}

\textit{Circular Rydberg atoms--} The atomic orbital of a circular Rydberg state with principal quantum number $n$ is a toroid centered at radius $r_{nc}=a_0 n^2$ (where $a_0$ is the Bohr radius) with a radial fuzziness $\Delta r_{nc}$ such that $\Delta r_{nc} /r_{nc}\ll1$ for $n\gg1$ \cite{haroche2006}. From this property follows that circular atoms have a long free-space  lifetime $\tau_{nc}\propto n^5$ (where $\tau_{50c} \approx 28$~ms) in non-demanding cryogenic conditions ($\lesssim$1~K). This timescale can be understood quantitatively in semiclassical terms: a charged electron orbiting at such distances from the nucleus will emit Larmor radiation at a photon rate $\approx1/\tau_{nc}$. The photon can be shown to have an energy $\hbar \omega_{nc}\propto n^{-3}$ \footnote{This can be shown by a semiclassical application of Kepler's third law for planetary motion \cite{haroche2006}.}, which reflects the dipole transition frequency of the single dominant decay channel of these states in free space $|nc\rangle \rightarrow |(n-1)c\rangle$ (where $\omega_{50c}/2\pi \approx 54$~GHz and $\omega_{100c}/2\pi \approx 6$~GHz, for example) \cite{haroche2006}. We estimate [from the pessimistic assumptions that i) all electron-photon scattering events are causing unwanted transitions and ii) using the Thomson cross section $\sigma_c=8 \pi r_e^2 / 3$ where $r_e$ is the `classical radius for the electron'] that the coherence lifetime in the thread trap is limited by photon scattering to tens of seconds \footnote{These considerations are specific to circular atoms and set them apart from low-angular momentum Rydberg atoms commonly used in state-of-the-art experiments. See \cite{Nguyen2018} for a related discussion.}.

The preparation \cite{Cantat2020} and laser trapping \cite{Cortinas2020} of cold circular atoms has been recently demonstrated experimentally, allowing validation, to a certain degree, of the expected theoretical properties. Using Rubidium and having no optical transitions available, ordinary laser dipole trapping was non-viable. Instead, a trap based on the ponderomotive potential \cite{landau1976,Dutta2000} was used. The principle of ponderomotive trapping of Rydberg atoms consists in trapping the dark-seeking Rydberg electron and, by doing so, trapping the entire atom bound to it. The trapping energy is then obtained by integrating the laser intensity over the electron wavefunction. The trap we propose here exploits this fact in the regime where the atomic orbital is larger than the laser intensity distribution at the focal point.

\textit{The thread trap--} In Figure \ref{fig:80c_trap} (a), we show a sketch of the atom threaded by a tightly focused Gaussian beam. In Figure \ref{fig:80c_trap} (b) and (c) we show the trapping potential $V$ created taking a laser at wavelength $\lambda = 420$~nm. The beam propagates along the $Z$-axis and we name the radial (cylindrical) coordinate $\rho$. Due to the subatomic-scale beam needed, the trap is tridimensional as the circular atom cannot run along the axis of the strongly divergent beam. The atom is thus localized within a fraction of the Rayleigh length along the axial direction. It is important to note that, since this trap does not rely on a dipole resonance, the wavelength $\lambda$ is comparatively unconstrained. Although the trapping potential depth scales as $\lambda^2$ for a given laser power (see below and \cite{landau1976}), the main consideration for implementation will likely be the diffraction-limited waist at $\approx\lambda$. Note that the freedom in laser wavelength, beam waist, and the principal quantum number $n$ provides flexibility and different configurations of interest. We illustrate this tunability in Figure \ref{fig:80c_trap} by plotting the trapping potential for $n=60,\;80$ and $100$. For $n=60$ we get the strongest linear confinement, which vanishes as $n$ grows for a fixed beam waist.

The circular wavefunction \footnote{A directing electric field is needed to impose a quantization axis. See \cite{Cortinas2020}} is found by solving the Schr\"odinger equation for the Hydrogen atom (an excellent approximation for circular Rydberg states of alkali metals), and reads in spherical coordinates \cite{haroche2006}

$$\psi_{nc}(r, \theta, \phi)=\frac{e^{-r / n a_0}}{\left(\pi a_0^3\right)^{1 / 2}} \frac{1}{n n !}\left(-\frac{r}{a_0 n} \sin \theta e^{i \phi}\right)^{n-1}  .$$

The integral $V\propto\int I|\psi_{nc}|^2 d\vec r$, over the tightly focused Gaussian beam of intensity $I$, is used to compute numerically the solid lines in Figure \ref{fig:80c_trap} (b) and (c). In the following, however, we approximate the circular orbital by a circular one-dimensional Bohr orbit to get analytical expressions for the properties of the trap. The confining potential computed under this semiclassical approximation is shown as dashed lines in Figure \ref{fig:80c_trap} (b) and (c). The approximation is remarkably good in the parameter regimes explored here. The ponderomotive potential for an atom whose nucleus is located at $(X,Y,Z)$ reads then

\begin{align*}
V(X,Y,Z) \approx \frac{1}{2 \pi} \int_0^{2 \pi}& d \theta \frac{q_e^2}{ m_e \epsilon_0 c} \frac{\lambda^2}{(2 \pi c)^2} \frac{\mathcal{P}_0}{\pi w^2(Z)}\\
&\times e^{-\frac{2}{w^2(Z)}\left[\left(X_n+X\right)^2+\left(Y_n+Y\right)^2\right]},
\end{align*}

where $q_e$ and $m_e$ are the charge and mass of the electron, $\epsilon_0$ is the permittivity of free space and $ c$ is the speed of light in vacuum, $\lambda$ is the wavelength of the threading laser, $\mathcal{P}_0$ is the laser power, $X_n=r_{nc} \cos \theta$ and $Y_n=r_{nc} \sin \theta$ are the circular Rydberg electron coordinates with respect to the atomic nucleus, $Z$ is the position of the nucleus (and of the electron) along the laser propagation axis, and $w(Z)=w_0 \sqrt{1+\left(Z / Z_R\right)^2}$ is the Gaussian waist of the laser beam, where $Z_R=\pi w_0^2/\lambda$ is the Rayleigh length of the beam. Here we have placed the origin of the coordinates at the focal point of the beam.

The mechanical trapping frequency along the axial direction for a circular Rydberg state $|nc\rangle$ at $\rho=\sqrt{X^2+Y^2} = 0$  is found by straightforward differentiation to be

$$
\omega_z^2=\frac{q_e^2}{m_e \epsilon_0 c} \frac{\lambda^2}{(2 \pi c)^2} \frac{2 \mathcal{P}_0}{\pi w_0^4} \frac{\left(2 r_{nc}^2-w_0^2\right) e^{-2 r_{nc}^2 / w_0^2}}{ Z_R^2  m_{\mathrm{Rb}}},
$$

where $m_{\mathrm{Rb}}$ is the mass of $^{87}$Rb. The axial nonlinearity from Taylor expanding the trapping potential around $Z=0$ as $V(\rho =0,Z) \approx k_{0,z}+k_{2,z} Z^2 + k_{4,z} Z^4$ is

$$k_{4,z}=\frac{q_e^2}{ m_e \epsilon_0 c} \frac{\lambda^2}{(2 \pi c)^2} \frac{\mathcal{P}_0}{\pi w_0^6}\frac{(2r_{nc}^4 - 4r_{nc}^2w_0^2 + w_0^4)e^{-2 r_{nc}^2 / w_0^2}}{Z_R^4},$$
while the associated Kerr coefficient \footnote{Which is nothing but the fourth order nonlinearity in units of frequency, obtained by measuring it in units of the zero point spread of the oscillator.}, defined as $K_z = \frac{3\hbar}{m_{\mathrm{Rb}}}\frac{k_{4,z}}{k_{2,z}}$, reads

$$K_z = 3\hbar\frac{(2r_{nc}^4 - 4r_{nc}^2w_0^2 + w_0^4)}{m_{\mathrm{Rb}}\left(2 r_{nc}^2-w_0^2\right)w_0^2 Z_R^2}.$$

Similarly, for the radial direction, we find a closed-form expression in terms of the modified Bessel function of the first kind and of order zero \cite{abramowitz1972abramowitz}. It is then easy to compute the radial small oscillation frequency and the Kerr coefficient to be

$$ \omega^2_{\rho} = \omega^2_{z} \frac{2 Z_R^2}{w_0^2},\;\;\;\;\;\;\;\;\; K_{\rho} =  K_{z} \frac{Z_R^2}{2w_0^2}.$$

Finally, note that the ponderomotive potential over the closed-shell core of alkali metals is negligible since it is several thousand times heavier than an electron ($V\propto 1/\textrm{mass}$).

\begin{figure}[t]
    \centering
    \includegraphics[]{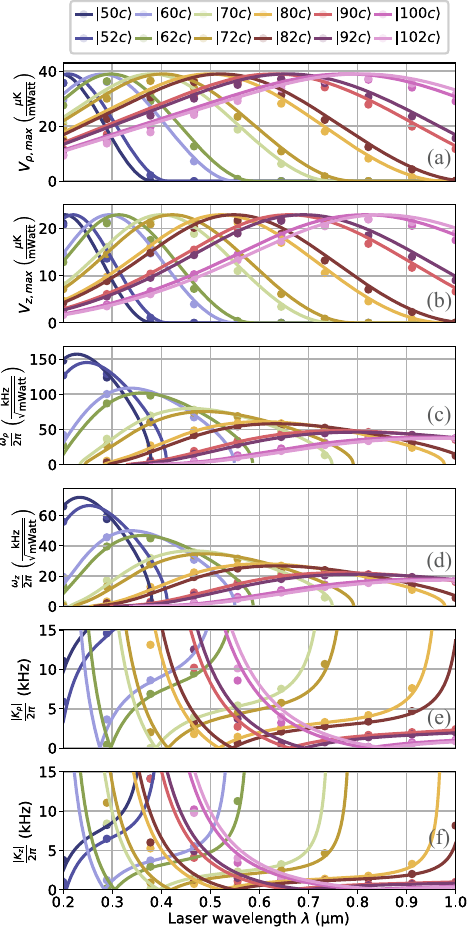}
    \caption{\textbf{Properties of the thread trap.} We show trap depths, trap frequencies, and Kerr coefficient in the thread trap configuration for a few $^{87}$Rb circular Rydberg states (colors) as a function of the trapping wavelength. These plots are made using a beam focused to a waist of $w_0=\lambda/2$. The dots correspond to the numerical integration over the toroidal wavefunction. The solid lines correspond to the semiclassical Bohr orbit approximation.}
    \label{fig:special}
\end{figure}

\textit{Properties of the thread trap--} In Figure \ref{fig:special} we show the trap depth, the small amplitude oscillation frequency and the Kerr nonlinearity of the trap as a function of the trapping wavelength $\lambda$. We plot these properties for both the radial and axial trapping axis, and for a few circular Rydberg states of $^{87}$Rb. As solid lines, we plot the semiclassical predictions, and the dots come from numerical integration of the laser intensity over $|\psi_{nc}|^2$.

Importantly, to produce this plot we have constrained the laser beam to a diffraction-limited Gaussian waist of

$$
w_0=\frac{\lambda}{\pi \mathrm{NA}} = \frac{\lambda}{2},
$$
using a numerical aperture $\mathrm{NA} \approx 0.65$. Note that $\mathrm{NA} \approx 0.7$ \cite{young2020} and even $\mathrm{NA} \approx 0.9$ \cite{robens2017} have been demonstrated in setups like the one required here. Requiring an attractive potential by $\left.\partial^2_{\rho} V\right|_{\rho,Z=0}>0$, the trapping condition simplifies to $\omega_z^2\propto (2 r_{nc}^2-w_0^2)>0$, and thus reads
$$
2 r_{nc}>\lambda / \sqrt{2},
$$
and it can be interpreted to mean that the Bohr atom needs to be large enough to fit a photon (of size $\approx \lambda/\sqrt{2}$) within its orbit. In Table~\ref{table:1} we write trapping conditions for a few Rydberg states. 

\begin{table}[h!]
\centering
\begin{tabular}{ccc} 
\hline \hline
$|n c\rangle$ & $\lambda$ \\
\hline
$|50 c\rangle$ & $<374 \mathrm{~nm}$ \\
$|60 c\rangle$ & $<538 \mathrm{~nm}$ \\
$|70 c\rangle$ & $<733 \mathrm{~nm}$ \\
$|80 c\rangle$ & $<957 \mathrm{~nm}$ \\
$|90 c\rangle$ & $<1212 \mathrm{~nm}$ \\
\hline \hline
\end{tabular}
\caption{\textbf{Trapping wavelengths.} Trapping conditions for beams focused to $w_0=\lambda/2$ and different circular states, as derived from $\lambda < 2 \sqrt{2} a_0 n^2$. Compare with Figure \ref{fig:special} (c) and (d) where the trapping frequency nulls.}
\label{table:1}
\end{table}

Note that one can relax this focusing condition by a non-diffraction-limited beam at a shorter wavelength. To give an example, the condition used here for a diffraction-limited laser at $\lambda =800$~nm can be achieved by a laser at $\lambda~=~$300~nm with a lens of NA $<$0.3.

In Figure \ref{fig:special} (a) and (b) we plot the radial and axial trap depth. We remark the high-trap depth that the thread trap provides. Comparable ponderomotive traps for Rydberg atoms \cite{Barredo2020} are more than five (and up to twenty) times less efficient. Their trap depth is limited by light ``holes'' providing typically barriers of $<4$~\textmu K/mWatt. Power inefficiency will be a bottleneck for proposals addressing tens of thousands or even millions of atoms in a Rydberg quantum computer \cite{bluvstein2021,Cong2022}, where Rydberg trapping will probably be a requirement \cite{Cohen2021}.

We see from Figure \ref{fig:special} (c) and (d) that essentially any pair of these circular Rydberg levels present a focusing diameter $\lambda$ at which the {trapping frequencies} are equal. This condition, for consecutive states $n$ and $n\pm1$ (or $n$ and $n\pm2$), happens near maximal trap depth. The condition along the radial direction coincides with the condition along the axial direction. This is a notable occurrence in the thread trap since it is a property of the convolution of the Gaussian beam with the circular orbit. Its origin is unrelated to the conditions  of equal polarizability found at particular wavelengths in ground-state dipole traps \cite{steck2007}.

In Figure \ref{fig:special} (e) and (f) the nonlinearities, as measured by the radial ($K_{\rho}$) and longitudinal ($K_{z}$) Kerr coefficient of the trap, are plotted and shown to cross for different circular states. The vertical plotting range of the Kerr coefficient is chosen to show the parameter regime where the perturbative approximation is meaningful while keeping the laser power at the few milliWatt level ($K_{\rho,z}\ll \omega_{\rho,z}$). Note that, since $K\propto k_{4}/k_{2}$, the Kerr coefficient is laser power independent. Another property, that can be read directly from Figure \ref{fig:special} (e) and (f), is that for essentially any pair of these circular Rydberg states, there exists a focusing diameter at which the nonlinearities are equal in magnitude (and opposite in sign). This condition happens close to maximum trap depth for neighboring states, too. This nonlinear parameter regime, provided sufficiently low dissipation conditions, will allow the observation of purely quantum dynamics and, in particular, the generation of Schr\"odinger Kerr-cats \cite{haroche2006} of a massive oscillator for the first time.

The last property we will mention is that the Kerr coefficient for any given state changes sign, passing through zero, (almost) simultaneously for the two trap axes ($K_{\rho}\approx K_z=0$). This differs from standard optical tweezer Gaussian traps or lattices which are constrained to have negative (softening) Kerr coefficient \footnote{The fundamental difference is that, here, the trapping potential does not follow from the shape of the beam only but it depends on the geometry of the wave function. The potential is given by the convolution between the two.}.  This happens near the maximum of trap depth too.

\textit{Discussion and concluding remarks--} We have proposed a trapping mechanism for Rydberg atoms that relies on the ponderomotive potential of the Rydberg electron and on the ability to thread a circular atomic orbital with a Gaussian beam. The trap exploits that the nontrivial shape and geometry of the Rydberg orbitals can be used as a tool and will provide a new handle on the atomic wavefunction.

Although threading an atomic orbital with light will be a challenge, it will be rewarding: 
freedom in the principal quantum number, beam waist, and the trapping wavelength allow the creation of deep, tunable traps. The trap can be used to trap ground states and Rydberg states alike and besides being power-efficient for Rydberg states, able to trap a cold Rydberg atom with as little as a milliWatt, the tunability of the energy landscape is interesting and may be useful for fundamental studies or provide a technical edge for specific applications.

We have discussed the tread trap mechanism for circular states of an alkali-metal atom but the idea can, in principle, be extended to non-circular and alkali-earth Rydberg states too \cite{teixeira2020, muni2022,Wilson2022}. With a simple Gaussian and lacking the symmetry of a circular state, low-angular momentum atoms will yield modest trap depth but one could think of using laser wavefronts other than a simple Gaussian to exploit the node location in the wavefunctions of different Rydberg states ($P$, $D$, etc.). For alkali-earth Rydberg atoms with two active electrons, one electron can be excited to a Rydberg state while the other is used to dipole-trap the core. This is the configuration of the experiments reported in \cite{Wilson2022}, were interesting questions about the competition between the dipolar and ponderomotive potentials arise. We note, however, that for alkali-earth atoms under both ponderomotive and dipole potentials the trap properties and the parameter regime available, as well as the trap-induced loss mechanisms (e.g. photoionization), and their Rydberg lifetime limitations (e.g. autoionization), are radically different from those expected for circular Rydberg Rubidium in a thread trap.

Finally, we specify a possible application that we believe is particularly promising. We point out that highly nonlinear driven systems are rich platforms for fundamental studies~\cite{Zurek1994,Habib1998}. In particular, we mention that the interplay of a Kerr nonlinearity and parametric squeezing in a massive quantum oscillator will bring new opportunities for cold atoms. Squeezing can be generated by exploiting the property that modulating the power of the trapping laser will modulate the trap frequency at constant Kerr coefficient. This nonlinear mechanical parametric oscillator could generate unprecedented control of quantum tunneling \cite{marthaler2006_}, interference in the classically forbidden region \cite{marthaler2007, Venkat2022} and bosonic encoding of quantum information \cite{cochrane1999,grimm2020,Frattini2022} in the mechanical degree of freedom of a neutral atom. Much like with interacting ions \cite{bruzewicz2019trapped}, the Rydberg interaction provides the opportunity to perform logical gates between two or more oscillators \cite{puri2019}. As such, this trap may bring the mechanical motion of cold circular Rydberg Rubidium atoms into a regime where only highly nonlinear, low-dissipation, superconducting quantum circuits operate today \cite{kirchmair2013,grimm2020}, and the experiment could conceivably be made at room temperature \cite{Meinert2020,Wu2023,hölzl2024longlived}.

\textit{Acknowledgments --} RGC thanks Jean-Michel Raimond, Michel Devoret, and Brice Ravon for critical remarks on an early version of this manuscript. RGC acknowledges discussions with W. D. Phillips, H. Wu, M. Brune, Y. Machu, M. Favier, C. Sayrin, S. Gleyzes, J. Thompson, V. Sivak, W. Adamczyk, S. Zacarias and P. and V. Kurilovich. Support from the Center for Quantum Dynamics on Modular Quantum Devices (NSF Grant No. CHE-2124511) and from the Yale Quantum Institute is gratefully acknowledged.

\bibliography{bib}
\end{document}